\documentstyle[12pt,axodraw]{article}
\begin{document}
\pagestyle{empty}
\quad\vskip 1.5cm
\centerline{ hep-ph/0004170 \hfill  CUMQ/HEP 112}
\vskip 1cm
\begin{center}{\bf \large
The Neutron Electric Dipole Moment and CP-violating Couplings \\
in the Supersymmetric Standard Model without $R$-parity}

\vskip 1cm

{\large
Darwin Chang$^{a}$, We-Fu Chang$^{a}$,
Mariana Frank$^{b}$,  and Wai-Yee Keung$^{c}$}\\

{\em $^a$NCTS and Physics Department, National Tsing-Hua University,\\
Hsinchu 30043, Taiwan, R.O.C.}\\
{\em $^b$Concordia University, Montreal, Quebec, Canada, H3G 1M8}\\
{\em $^c$Physics Department, University of Illinois at Chicago, IL
     60607-7059,  USA}\\

\end{center}

\date{\today}
\vskip 1.5cm
\begin{abstract}
We analyze the neutron electric dipole moment (EDM) in the Minimal
Supersymmetric Model with explicit $R$-parity violating terms. The leading
contribution to the EDM occurs at the 2-loop level and is
dominated by the chromoelectric dipole moments of quarks,
assuming there is no tree-level mixings between
sleptons and Higgs bosons or between leptons  and gauginos.
Based on the experimental constraint on the neutron EDM, we set limits
on the imaginary parts of  complex couplings ${\lambda'}_{ijk}$ and
${\lambda}_{ijk}$ due to the virtual $b$-loop or $\tau$-loop.
\end{abstract}

%
%
\newpage
%
\pagestyle{plain}

The minimal supersymmetric standard model (MSSM)~\cite{one}
has been widely considered as
a leading candidate for new physics beyond Standard Model.
However, unlike the standard model, supersymmetry allows
renormalizable interactions  which break $R$-parity defined as
$(-1)^{3B+L+F}$ and violate the lepton and/or the baryon numbers.
It is in fact one of the main theoretical weaknesses of these models
because $R$-parity conservation  is an ad hoc imposition which may
or may not have a fundamental theoretical basis.
Therefore, instead of neglecting them completely, it is interesting
to ask how small  these $R$-parity breaking
($\not\! R$) couplings could be by
investigating directly phenomenological constraints\cite{two}.

The most general renormalizable $R$-violating superpotential using only
the MSSM superfields is
\begin{eqnarray}
W_{\not\! R} = \lambda_{ijk}L_i L_j E_k^C +
\lambda^{\prime}_{ijk}L_i Q_j D_k^c +
\lambda_{ijk}^{\prime\prime}U_i^c D_j^c D_k^c +\mu_j L_j H_2 \ .
\end{eqnarray}
Here, $i,j,k$ are generation indices.
The couplings $\lambda^k_{ij}$  and $\lambda^{k \prime \prime}_{ij}$ must
be antisymmetric in flavor,
$ \lambda_{ijk}= -\lambda_{jik}$ and
      $ \lambda^{\prime\prime}_{ijk}=- \lambda^{\prime\prime}_{ikj}$.
There are 36 lepton number non-conserving couplings (9 of the
$\lambda$ type and 27 of the $\lambda'$ type) and 9 baryon number
non-conserving couplings (all of the $\lambda{''}$ type) in
Eq.(1). To avoid rapid proton decay, it is usually assumed in the
literature that $\lambda$, $\lambda'$ type couplings do not
coexist with $\lambda{''}$ type couplings. This can be achieved
easily by imposing baryon number symmetry.  The bilinear terms
$\mu_j L_j H_2$ contribute to lepton flavor and number violation
and could be responsible for neutrino masses. Phenomenologically,
many of these couplings have been severely constrained using
low-energy processes or using high energy data at the colliders
\cite{dreiner}--\cite{cern}. In this paper, we shall not consider
$\lambda_{ijk}^{\prime\prime}$ and $\mu_j$ couplings.

However, most of the bounds in the literature constrain the real part of
the trilinear couplings, or the product of trilinear couplings. The
exception is the bound coming
from the $\epsilon_K$ which constrains $\mbox{Im} (\lambda^{\prime}_{i12}
\lambda^{\prime \ast}_{i21})< 8 \times 10^{-12}$ \cite{bsb}. We propose to
study the neutron electric dipole moment, which is
    tightly bound by experiment, and thus obtain limits on the imaginary
parts of different products of trilinear couplings from the ones imposed
by $\epsilon_K$.

The electric dipole moment of an elementary fermion is defined through its
electromagnetic form factor $F_3(q^2)$ in the (current) matrix
element:
\begin{eqnarray}
\label{formfactors}
\langle f(p')|J_{\mu}(0)|f(p) \rangle=\bar{u}(p')\Gamma_{\mu}(q)u(p),
\end{eqnarray}
where $q=p'-p$ and
\begin{eqnarray}
\label{current}
\Gamma_{\mu}(q)=F_1(q^2)\gamma_{\mu}
+{F_2(q^2)\over 2m} i\sigma_{\mu\nu}q^{\nu}
+F_A(q^2)
(\gamma_{\mu}\gamma_5q^2-2m\gamma_5q_{\mu})
+{F_3(q^2)\over 2m} \sigma_{\mu\nu}
\gamma_5q^{\nu},
\end{eqnarray}
with $m$ the mass of the fermion
and $F_1(0)=e_f$. The electric dipole moment
(EDM) of the fermion field $f$ is then given by
\begin{eqnarray}
\label{edm}
d_f={eF_3(0)\over 2m}\ ,
\end{eqnarray}
corresponding to the effective dipole interaction
\begin{eqnarray}
\label{dipole}
{\cal L}_{\rm EDM}=
- \frac{i}{2} d_f \bar{f}\sigma_{\mu\nu}\gamma_5 f F^{\mu\nu} \ .
\end{eqnarray}
In the static limit this corresponds to an effective Lagrangian
${\cal L}_{\rm EDM} \to
d_f \Psi^{+}_A {\vec \sigma}\cdot{\vec E} \Psi_A$,
where $\Psi_A$ is the large
component of
the Dirac field. Similarly the quark chromoelectric dipole moment
(CEDM) is
the coefficient $d_q^g$ in the effective operator:
\begin{eqnarray}
\label{cedm}
{\cal L}_{\rm CEDM}= - \frac{i}{2} d_q^g \bar{q}\sigma_{\mu\nu}\gamma_5
{\lambda^a\over2}q G^{a \mu \nu} \quad .
\end{eqnarray}

The relevant Lagrangian for generating an EDM is:
\begin{eqnarray}
{\cal L}
= -\left(\frac12 \sum_{ij}
\Psi_i \frac{\partial^2 W}{\partial \phi_i \partial \phi_j}\Psi_j
+\hbox{ H.c.}\right)+\cdots
=\left[
\lambda^{\prime}_{ijk} (-\tilde{\ell_i} u_j d_k^c
                          +\tilde{\nu_i } d_j d_k^c) + \hbox{ H.c.}\right]
      + \cdots \quad.
\end{eqnarray}

It has been shown \cite{godbole} that there is no one-loop
contributions to EDMs based on $\lambda$, $\lambda'$ or $
\lambda^{\prime\prime}$ couplings based on helicity properties and
symmetry. Here we briefly review its origin. It is easy to show
that one cannot induce EDMs from the diagram that requires the
external mass insertion due to the equation of motion. As a result
of this lemma, proper helicities for external fermion lines have
to come directly from vertices. Let us look at the electron EDM,
which needs external $L$ and $E^c$. For the correct quantum
number, possible one loop contributions have to be  proportional
to either (1)$\lambda \lambda^*$ or (2)$\lambda' \lambda'^*$.
Based on the above lemma, the external $L$ and $E^c$ are required
to come directly from vertices. Case (1) cannot produce the
helicity flip. Case (2) is even worse, there is no vertex to give
$E^c$. So the one-loop electron EDM is absent.
For the $d$ quark EDM,
possibilities are either
(1) $\lambda' \lambda'^*$ or
(2) $\lambda''\lambda''^*$.
Case (1) does not work because both $d_L$ and $d^c$ have to come from
a $CP$-even product of a complex conjugated pair of  vertices, and
case (2) fails badly because there is no vertex to give an external $d_L$.
Similar reason follows for the $u$ quark EDM.
As a reminder, there are one-loop EDM amplitudes \cite{refchoi}
related to the bilinear term
$\mu_j L_j H_2$, which mixes sleptons and Higgs bosons etc.
We do not consider these couplings $\mu_i$ in this work.

At the two-loop level, a number of different types of
configurations contribute, which we classify as
rainbow-like (I), overlapping (II), tent-like (III) and Barr-Zee
(IV)\cite{barr-zee} types. The rainbow-like graphs (I) are those
with two concentric boson loops, the outer of which must be a
charged Higgs loop (for the same reason that 1-loop graphs do not
exist). The inner loop may be a left or right sfermion. The
complete set of this type of graphs is given in Fig.~1a.
The complete set of overlapping type of graphs is given in Fig.~1b.
In this case, one of them must be a charged Higgs, the other a left or
right sfermion.
\begin{center}
\begin{picture}(200,125)(0,0)
 \ArrowLine(10,15)(40,15)   \Text(25,7)[c]{$d_L$}
 \ArrowLine(40,15)(57,15)  \Text(45,7)[c]{$u_{jR}$}
 \Line(55,17)(60,13) \Line(55,13)(60,17)
 \ArrowLine(57,15)(75,15) \Text(63,7)[c]{$u_{jL}$}
      \BCirc(75,15){3}
 \ArrowLine(75,15)(125,15) \Text(100,7)[c]{$d_{kR}$}
      \GCirc(125,15){3}{0.5}
 \ArrowLine(125,15)(160,15) \Text(142,7)[c]{$u_{mL}$}
 \ArrowLine(160,15)(190,15) \Text(175,7)[c]{$d_R$}

 \DashArrowArcn(100,15)(60,180,90){3}  \Text(40,60)[c]{$H^{-}$}
 \DashArrowArcn(100,15)(60,90,0){3}    \Text(160,60)[c]{$H^{-}$}

 \DashArrowArc(100,15)(25,0,180){3}  \Text(95,55)[c]{$\tilde{l}_{iL}$}

 \Photon(100,115)(100,75){2}{6} \Text(112,95)[c]{$\gamma$}

 \end{picture}

\end{center}
Fig.~1a  (i)  Rainbow-like  diagram for the $d$ quark. The generic
$\not R$ vertex is marked by $\circ$  and its complex conjugate by
$\bullet$.
\newpage
\begin{center}
\begin{picture}(200,125)(0,0)
 \ArrowLine(10,15)(40,15)   \Text(25,7)[c]{$u_L$}
 \ArrowLine(40,15)(75,15)  \Text(53,7)[c]{$d_{mR}$}
      \GCirc(75,15){3}{0.5}
 \ArrowLine(75,15)(125,15) \Text(100,7)[c]{$u_{jL}$}
      \BCirc(125,15){3}
 \ArrowLine(125,15)(142,15) \Text(133,7)[c]{$d_{kR}$}
 \Line(140,17)(144,13) \Line(140,13)(144,17)
 \ArrowLine(142,15)(160,15) \Text(150,7)[c]{$d_{kL}$}
 \ArrowLine(160,15)(190,15) \Text(175,7)[c]{$u_R$}

 \DashArrowArc(100,15)(60,90,180){3}  \Text(40,60)[c]{$H^{-}$}
 \DashArrowArc(100,15)(60,0,90){3}    \Text(160,60)[c]{$H^{-}$}

 \DashArrowArcn(100,15)(25,180,0){3}  \Text(95,55)[c]{$\tilde{l}_{iL}$}

 \Photon(100,115)(100,75){2}{6} \Text(112,95)[c]{$\gamma$}

 \end{picture}
 \\

\begin{picture}(200,125)(0,0)
 \ArrowLine(10,15)(40,15)   \Text(25,7)[c]{$u_L$}
 \ArrowLine(40,15)(75,15)  \Text(53,7)[c]{$d_{mR}$}
      \GCirc(75,15){3}{0.5}
 \ArrowLine(75,15)(125,15) \Text(100,7)[c]{$d_{jL}$}
      \BCirc(125,15){3}
 \ArrowLine(125,15)(142,15) \Text(133,7)[c]{$d_{kR}$}
 \Line(140,17)(144,13) \Line(140,13)(144,17)
 \ArrowLine(142,15)(160,15) \Text(150,7)[c]{$d_{kL}$}
 \ArrowLine(160,15)(190,15) \Text(175,7)[c]{$u_R$}

 \DashArrowArc(100,15)(60,90,180){3}  \Text(40,60)[c]{$H^{-}$}
 \DashArrowArc(100,15)(60,0,90){3}    \Text(160,60)[c]{$H^{-}$}

 \DashArrowArcn(100,15)(25,180,0){3}  \Text(95,55)[c]{$\tilde{\nu}_{iL}$}

 \Photon(100,115)(100,75){2}{6} \Text(112,95)[c]{$\gamma$}
 \end{picture}
 \\
Fig.~1a  (ii)   Rainbow-like  diagram for the $u$ quark.
\end{center}

\begin{center}
\begin{picture}(250,125)(0,0)
 \ArrowLine(10,25)(37,25)   \Text(25,12)[c]{$d_R$}
 \ArrowLine(37,25)(87,25)  \Text(60,12)[c]{$u_{Lj}$}
    \BCirc(87,25){3}
  \ArrowLine(125,25)(87,25) \Text(105,12)[c]{$l_{Li}$}
   \Line(123,27)(127,23) \Line(123,23)(127,27)
 \ArrowLine(162,25)(125,25) \Text(145,12)[c]{$l_{Ri}$}
 \ArrowLine(212,25)(162,25) \Text(175,12)[c]{$\nu_{Li}$}
     \GCirc(212,25){3}{0.5}
 \ArrowLine(212,25)(240,25) \Text(225,12)[c]{$d_L$}

\DashCArc(100,25)(62,0,180){3} \Text(95,100)[c]{$H^{-}$}

 \DashCArc(150,25)(62,90,180){3}
 \DashArrowArcn(150,25)(62,90,0){3}    \Text(185,90)[c]{$\tilde{d}_{Rk}$}

\Photon(150,87)(150,125){5}{3}  \Text(160,110)[c]{$\gamma$}
 \end{picture}
 \\
Fig.~1b (i) Overlapping  diagram for the $d$ quark using
$\lambda'$.
\end{center}

\begin{center}
\begin{picture}(250,125)(0,0)
 \ArrowLine(10,25)(37,25)   \Text(25,12)[c]{$u_R$}
 \ArrowLine(37,25)(87,25)  \Text(60,12)[c]{$d_{Lj}$}
      \BCirc(87,25){3}
  \ArrowLine(162,25)(87,25) \Text(105,12)[c]{$\nu_{Li}$}
 \ArrowLine(187,25)(162,25) \Text(175,12)[c]{$l_{Ri}$}
   \Line(185,27)(189,23) \Line(185,23)(189,27)
 \ArrowLine(212,25)(187,25) \Text(199,12)[c]{$l_{Li}$}
       \GCirc(212,25){3}{0.5}
 \ArrowLine(212,25)(240,25) \Text(225,12)[c]{$u_L$}
\DashCArc(100,25)(62,0,180){3} \Text(95,100)[c]{$H^{-}$}

 \DashCArc(150,25)(62,90,180){3}
 \DashArrowArcn(150,25)(62,90,0){3}    \Text(185,90)[c]{$\tilde{d}_{Rk}$}

\Photon(150,87)(150,125){5}{3}  \Text(160,110)[c]{$\gamma$}
 \end{picture}
 \\
Fig.~1b (ii) Overlapping diagram for the $u$ quark using
$\lambda'$.
\end{center}

\begin{center}
\begin{picture}(225,137)(0,0)
 \ArrowLine(10,62)(50,62)   \Text(25,50)[c]{$d_L$}
    \BCirc(50,62){3}
 \ArrowLine(50,62)(112,62)  \Text(75,50)[c]{$d_{kR}$}
    \GCirc(112,62){3}{0.5}
 \ArrowLine(112,62)(175,62)  \Text(137,50)[c]{$u_{jL}$}
 \ArrowLine(175,62)(215,62)  \Text(187,50)[c]{$d_R$}

 \DashArrowLine(112,62)(112,95){3} \Text(125,77)[c]{$\tilde{l}_{iL}$}
  \Line(110,93)(114,97) \Line(110,97)(114,93)
 \DashArrowLine(112,95)(112,125){3} \Text(125,112)[c]{$\tilde{l}_{iR}$}

 \DashArrowArc(112,62)(62,90,180){5}  \Text(50,105)[c]{$\tilde{\nu}_{iL}$}

 \DashArrowArcn(112,62)(62,90,0){5}    \Text(175,105)[c]{$H^{-}$}

\Photon(155,62)(155,5){2}{5} \Text(165,25)[c]{$\gamma$}
 \end{picture}
 \\

 \begin{picture}(225,137)(0,0)
 \ArrowLine(10,62)(50,62)   \Text(25,50)[c]{$d_L$}
    \BCirc(50,62){3}
 \ArrowLine(50,62)(112,62)  \Text(75,50)[c]{$d_{kR}$}
    \GCirc(112,62){3}{0.5}
 \ArrowLine(112,62)(175,62)  \Text(137,50)[c]{$u_{jL}$}
 \ArrowLine(175,62)(215,62)  \Text(187,50)[c]{$d_R$}

 \DashArrowLine(112,62)(112,125){3} \Text(125,90)[c]{$\tilde{l}_{iL}$}

 \DashArrowArc(112,62)(62,90,135){5}  \Text(75,125)[c]{$\tilde{\nu}_{iR}$}
 \Line(66,106)(72,106) \Line(69,109)(69,103)
 \DashArrowArc(112,62)(62,135,180){5}  \Text(40,95)[c]{$\tilde{\nu}_{iL}$}

 \DashArrowArcn(112,62)(62,90,0){5}    \Text(175,105)[c]{$H^{-}$}

\Photon(155,62)(155,5){2}{5} \Text(165,25)[c]{$\gamma$}
 \end{picture}
 \\
Fig.~1c (i) Tent-like  diagram for the $d$ quark using $\lambda'$.
\end{center}

\begin{center}
\begin{picture}(225,137)(0,0)
 \ArrowLine(10,62)(50,62)   \Text(25,50)[c]{$u_L$}
    \BCirc(50,62){3}
 \ArrowLine(50,62)(112,62)  \Text(75,50)[c]{$d_{kR}$}
    \GCirc(112,62){3}{0.5}
 \ArrowLine(112,62)(175,62)  \Text(137,50)[c]{$d_{jL}$}
 \ArrowLine(175,62)(215,62)  \Text(187,50)[c]{$u_R$}

 \DashArrowLine(112,62)(112,125){3} \Text(125,90)[c]{$\tilde{\nu}_{iL}$}


 \DashArrowArc(112,62)(62,90,135){5} \Text(75,125)[c]{$\tilde{l}_{iR}$}
  \Line(66,106)(72,106) \Line(69,109)(69,103)
 \DashArrowArc(112,62)(62,135,180){5}  \Text(40,95)[c]{$\tilde{l}_{iL}$}

 \DashArrowArc(112,62)(62,0,90){5}    \Text(175,105)[c]{$H^{-}$}

\Photon(155,62)(155,5){2}{5} \Text(165,25)[c]{$\gamma$}
 \end{picture}
 \\
Fig.~1c (ii) Tent-like  diagram for the $u$ quark using
$\lambda'$.
\end{center}

\begin{center}
\begin{picture}(200,150)(0,0)
 \ArrowLine(180,40)(160,40)   \Text(170,30)[c]{$u_R$}
 \ArrowLine(160,40)(100,40)  \Text(120,30)[c]{$d_{kL}$}
 \Line(98,38)(102,42) \Line(98,42)(102,38)
 \ArrowLine(100,40)(40,40) \Text(80,30)[c]{$d_{kR}$}
 \ArrowLine(40,40)(20,40) \Text(30,30)[c]{$u_L$}

 \DashArrowLine(40,40)(73,73){3}    \Text(40,60)[c]{$\tilde{l}_{iL}$}
 \DashArrowLine(128,73)(160,40){3}  \Text(160,60)[c]{$H^{-}$}

 \Oval(100,90)(28,35)(0)
 \Photon(100,120)(100,150){3}{4} \Text(110,135)[c]{$\gamma$}

 \Text(110,55)[c]{$u_{jL}$}
 \LongArrow(102,62)(98,62)
 \Text(50,100)[c]{$d_{mR}$}
 \LongArrow(65,88)(65,92)
 \BCirc(73,73){3}
 \GCirc(40,40){3}{0.5}
 \end{picture}
 \\
Fig.~1d Barr-Zee type graph for $u$ quark EDM
\end{center}
\newpage

The tent-like graphs (III) have a trilinear bosonic vertex.
Again, the 3 different
boson legs can be two sfermions and one charged Higgs (in all
possible configurations).  The complete set of this type of graphs
is given in Fig.~1c.  Careful consideration of all the type I--III
graphs shows that their contributions are suppressed by both one power of
light quark mass plus some CKM mixing angles factor compared to those of
type IV (Fig.~2). Therefore, one expects the Barr-Zee type of
contributions dominate and we shall study them in detail next.

\begin{center}
\begin{picture}(200,160)(0,0)
\Oval(100,80)(30,50)(0) \Gluon(100,110)(100,160){5}{5}
\Text(110,150)[lc]{$\downarrow$ gluon, $(k,\mu)$}
\DashArrowLine(60,60)(20,20){5} \Gluon(140,60)(180,20){5}{5}
\Text(35,40)[rc]{$\tilde{\nu_L}(-p)$}
\Line(97,47)(103,53) \Line(97,53)(103,47)
\Text(80,42)[cc]{${b_R}^c$} 
\Text(120,65)[cc]{$l$}
\Text(140,90)[rc]{$l+q$} \LongArrow(135,102)(140,98)
\Text(70,115)[rc]{$b_L$} \Text(60,90)[lc]{$l+p$}
\LongArrow(60,98)(65,102 ) 
\LongArrow(80,53)(85,51 )  
\LongArrow(121,53 )(117,51)  
\LongArrow(160,55)(170,45) 
\Text(175,50)[lc]{gluon $(q,\nu)$}
\ArrowLine(0,20)(20,20) \Text(10,10)[cc]{${d_R}^c$}
\ArrowLine(180,20)(20,20) \Text(100,10)[cc]{$d_L(q)$}
\ArrowLine(200,20)(180,20) \Text(190,10)[cc]{$d_L$}
\end{picture}
\end{center}
Fig.~2\quad
A typical two-loop diagram of the Barr-Zee type.
Note that there are 3 ways to insert mass.

In Ref. \cite{godbole}, only a rough estimate of the two-loop
contributions to EDMs is provided.  We shall present here a
complete calculation of the quark (or electron)  EDM and the quark
CEDM at the two-loop level  due to the Barr-Zee type mechanism and
show that the neutron EDM is dominated by the CEDM of the $d$ quark.
This calculation leads to more stringent bounds than previously
obtained.

There is another class of Barr-Zee graphs with sneutrino line
replaced by the charged slepton line and corresponding modifications of the
fermions charges in the loop.  The calculations of these type of
graphs are very similar to the one in the charged Higgs models of
CP violation as in Ref.\cite{bck,cck}.  Comparing the charged
Higgs contributions to the EDM in Ref.\cite{bck,cck} with the neutral Higgs
contributions given in Ref.\cite{barr-zee}, one can observe that
the neutral Higgs contributions generally dominates given
comparable coupling constants and boson masses.  Therefore, we
shall only give details of sneutrino contributions here.

\section*{$\tilde{\nu}gg$ vertex of the inner loop}
The two-loop diagram to the CEDM of  the $d$-type quark can appear
with the coupling $ \lambda^{\prime}_{ijk}$ through the virtual
vertex $\tilde\nu gg$. The amplitude of the inner loop in terms of
the leading gauge invariant terms is:

\begin{equation}
\Gamma^{\mu\nu}= S(q^2)[k^\nu q^\mu -k\cdot q g^{\mu\nu}]
                    +P(q^2)[i \epsilon^{\mu\nu\alpha\beta} p_\alpha
                    q_\beta ]
\ ,\end{equation}
where $S$ and $P$ correspond to  scalar and pseudoscalar form factors
respectively:

\begin{eqnarray}
S(q^2) &=&
{ m_b g_s^2  \lambda^{\prime *}_{i33}\over 16\pi^2}
      \int^1_0 dy {1-2y(1-y) \over m_b^2-y(1-y)q^2 }   \ ,\nonumber\\
P(q^2)&=&
{  m_b g_s^2 \lambda^{\prime *}_{i33}\over 16 \pi^2}
      \int^1_0 dy {1\over m_b^2-y(1-y)q^2 }\ .\nonumber\
\end{eqnarray}

\section*{Second loop}

Combining the two twisted diagrams and the two choices of
sneutrino flow directions, we have a combinatoric factor of 4 in the
the two loop CEDM amplitude.
In the convention of Eqs.(2--6),
we  obtain   the
CEDM of the $d$ quark at the scale of $m_b$,
\begin{eqnarray}
\left({d_d^g \over g_s}\right)_{m_b} &=& { m_b g_s^2|_{m_b} \over 128 \pi^4}
\mbox{Im}( \lambda^{\prime *}_{i33}\lambda^{\prime}_{i11})
\int^1_0 dy \int^\infty_0
{ Q^2 d(Q^2) ( 1-y(1-y))
\over( m_b^2+y(1-y)Q^2)(Q^2+M^2_{\tilde{\nu_i}})Q^2  }\nonumber\\
&=& { \alpha_s  \mbox{Im}( \lambda^{\prime * }_{i33}
\lambda^{\prime}_{i11} )
     \over 32 \pi^3}{m_b \over  M_{\tilde \nu_i}^2   }
     \cdot F\left({m_b^2\over M_{\tilde \nu_i}^2}\right) \quad ,
\label{eq:cedm}\end{eqnarray}
with the loop function
\begin{eqnarray}
F(\tau)&=&\int^1_0 dy {(y^2-y+1)\over y(1-y)-\tau  }
\ln \left(\frac{y(1-y)}{\tau} \right)  \\
&\to&  {\pi^2\over3}+2+\ln\tau+(\ln\tau)^2 \quad \hbox{for } \tau\to 0
\ .\end{eqnarray}
Implicit sum over sneutrino flavors $i$ is assumed in the above.  The last
asymptotic form is useful because the ratio
$\tau=m_b^2/M^2_{\tilde{\nu_i}}$ is small. The large logarithmic
factor helps place a strong constraint on $\lambda'$ couplings.
Note that sneutrino is the heaviest particle in the loop.  At $m_b$ scale,
the sneutrino induces a four fermion interaction of $b$ and $d$ quarks.
As a result, by simple power counting, the gluonic loop is logarithmically
divergent which explains the large logarithmic  enhancement factor.

Replacing the gluon line by the photon line, we obtain the EDM of
the quark simply by substituting the color factor and the charge
factor.
\begin{equation}
\left({d_d^\gamma \over e}\right)_{m_b} = 6e_d e_b^2
\left({\alpha_{\rm em}\over \alpha_s}\right)_{m_b}
\left({d_d^g \over g_s}\right)_{m_b}
\ . \end{equation}

Now we address the QCD evolution of these Wilson's coefficients.
As an effective theory, the 4-fermion vertices of the form $(\bar
b b)(\bar d d)$  arise first in the energy scale below $M_{\tilde
\nu_i}$ when $\tilde\nu$ are integrated out of the theory. Then
the EDM and the CEDM of the $d$ quark arise below $m_b$ scale.
Therefore, the $\lambda'$ couplings in the above equations are
evaluated at the $m_b$ scale. We ignore the dressing of these
4-fermion vertices because of the small value of their couplings
and the slow running of $\alpha_s$ at such high energy scale. In
this perspective, the $\lambda^{\prime}$ factors in
Eq.~(\ref{eq:cedm}) are  defined at the short distance scale near
$M_{\tilde \nu}$.
Below $m_b$, the CEDM and the EDM of light quarks appear and they
evolve down to the hadronic scale $\Lambda_H$ by
\begin{equation}
\left({d_d^g \over  g_s}\right)_{\Lambda_H}
\left/\left({d_d^g\over g_s}\right)_{m_b}\right.
=         \left({g_s(m_b)\over g_s(m_c)}\right)^{\frac{4}{25}}
             \left({g_s(m_c)\over g_s(\Lambda_H)}\right)^{\frac{4}{27}}
=Z^g  \ ,
\end{equation}
\begin{equation}
\left({d_d^\gamma \over  e}\right)_{\Lambda_H}\left/
\left({d_d^\gamma\over e}\right)_{m_b}        \right.
=         \left({g_s(m_b)\over g_s(m_c)}\right)^{\frac{8}{25}}
             \left({g_s(m_c)\over g_s(\Lambda_H)}\right)^{\frac{8}{27}}
=Z^\gamma \ .
\end{equation}
Note that in some references \cite{cky}, a light quark mass coefficient
has been factored out so that the form of evolution equation looks
different from above.
We denote by $D_n^g$ ($D_n^\gamma$) the neutron EDM due to the
CEDM (EDM) of light quarks.  The $SU(6)$ relation gives:
\begin{equation}
\frac{D_n^g}{e}=
       \left[ \frac49\left(\frac{d^g_d}{g_s}\right)_{\Lambda_H}
             +\frac29\left(\frac{d^g_u}{g_s}\right)_{\Lambda_H}\right]
\ .\qquad
\frac{D_n^\gamma}{e}=
       \left[ \frac43\left(\frac{d^\gamma_d}{e}\right)_{\Lambda_H}
             -\frac13\left(\frac{d^\gamma_u}{e}\right)_{\Lambda_H}\right].
\end{equation}
For  $\alpha_s(M_Z)=0.12$ and
$g_s(\Lambda_H)/(4\pi)=1/\sqrt{6}$, the QCD evolution factors
$Z^\gamma$ and $Z^g$
are about 0.71 and 0.84 respectively.
Our formulas and numerical values are consistent with those in
Ref. \cite{cky} but differ from those in Ref. \cite{arnowitt}.

For completeness, we add another large contribution
to the $d$ quark EDM due to the $\tau$-lepton replacing the $b$ quark
inside the first loop.  We obtain two  independent contributions as:

\begin{equation}
\left({d_d^\gamma \over e}\right)_{m_b} \hbox{($b$-loop)}
    = -{ \alpha_{\rm em}   \over 16 \pi^3}
      \sum_{i=1,2,3} {3 e_d e_b^2 m_b \over  M_{\tilde \nu_i}^2   }
      \mbox{Im}(\lambda^{\prime * }_{i33}\lambda^{\prime}_{i11}
)
     \cdot F\left({m_b^2\over M_{\tilde \nu_i}^2}\right) \ ,
\end{equation}
\begin{equation}
\left({d_d^\gamma \over e}\right)_{m_\tau} \hbox{($\tau$-loop)}
    = -{ \alpha_{\rm em}   \over 16 \pi^3}
     \sum_{i\ne 3} {e_d m_\tau \over  M_{\tilde \nu_i}^2   }
     \mbox{Im}(\lambda^{*}_{i33}\lambda^{\prime}_{i11} )
     \cdot F\left({m_\tau^2\over M_{\tilde \nu_i}^2}\right)
    \quad .
\label{eq:dedm}\end{equation} The latter contribution from the
$\tau$-loop is induced at the $m_\tau$ scale and we need to
adjust the minor change in the QCD evolution. There are also
other Barr-Zee type diagrams from the exchange of $W^\pm$ or $Z$
gauge bosons. However, they are known to be giving smaller
contributions and thus we ignore them in our numerical study
\cite{bck,cck,pil}.

As $u_R$ is not directly involved in the $\not\! R$ interaction,
the $u$ quark  CEDM do not appear through $\lambda'$
in the form of Fig.~2.
Nonetheless, there are two-loop diagrams
Fig.~1a (ii), 1b(ii), 1c(ii) and 1d  which are suppressed by the
light quark mass and mixing angles.
Therefore, the  $\not\! R$ contribution to the neutron EDM is
dominated by the $d$ qaurk CEDM and EDM.
Assuming all $M_{\tilde{\nu_i}}$ are equal and taking typical
values, $M_{\tilde{\nu_i}}\approx 300$ GeV and $m_b\approx 4.5$
GeV, we have
\begin{equation}
D_n^g \simeq  5.46\times 10^{-21}    \  (e\mbox{-cm})\times  \sum_i
\mbox{Im}( \lambda^{\prime * }_{i33} \lambda^{\prime}_{i11} )
    \ ,
\end{equation}
\begin{eqnarray}
D_n^\gamma \simeq &&-1.03\times 10^{-22}    \ (e\mbox{-cm})   \times \sum_i
\mbox{Im}( \lambda^{\prime * }_{i33} \lambda^{\prime}_{i11} )  \nonumber\\
   && -1.92 \times 10^{-22}    \ (e\mbox{-cm})   \times \sum_{i\ne3}
\mbox{Im}( \lambda^{*}_{i33} \lambda^{\prime}_{i11} )
\ .
\end{eqnarray}
Our numerical result shows that the strongest constraint comes from
the CEDM of the $d$ quark.
Using the up-to-dated  experimental \cite{neutronedm} bound
$|D_n|<6.3\times 10^{-26}$ $e$-cm
and barring accidental cancellation among contributions,
we derive the  constraints:
\begin{eqnarray}
\sum_i \mbox{Im}(\lambda^{\prime * }_{i33} \lambda^{\prime}_{i11})
          &<&      1.2\times 10^{-5} \ ,  \\
\sum_{i\ne 3} \mbox{Im}(\lambda^{*}_{i33} \lambda^{\prime}_{i11})
          &<&      33\times 10^{-5} \ , \end{eqnarray}
for $M_{\tilde\nu}=300$ GeV.

In Fig.~3 we plot   both the photon
and gluon contributions to the neutron EDM versus the sneutrino mass
$M_{\tilde \nu}$ in the region of interest ($100$ to $600$ GeV)
with
$\sum_i \mbox{Im}(\lambda^{\prime * }_{i33} \lambda^{\prime}_{i11})$
or
$\sum_i \mbox{Im}(\lambda^{*}_{i33} \lambda^{\prime}_{i11})$
scaled  to $10^{-5}$.
One could see that the gluon contribution
consistently dominates the corresponding photon one by at least an order
of magnitude over the whole parameter space explored.
\vskip .5 cm
\begin{center}
\begin{picture}(400,240)(0,0)
\includegraphics{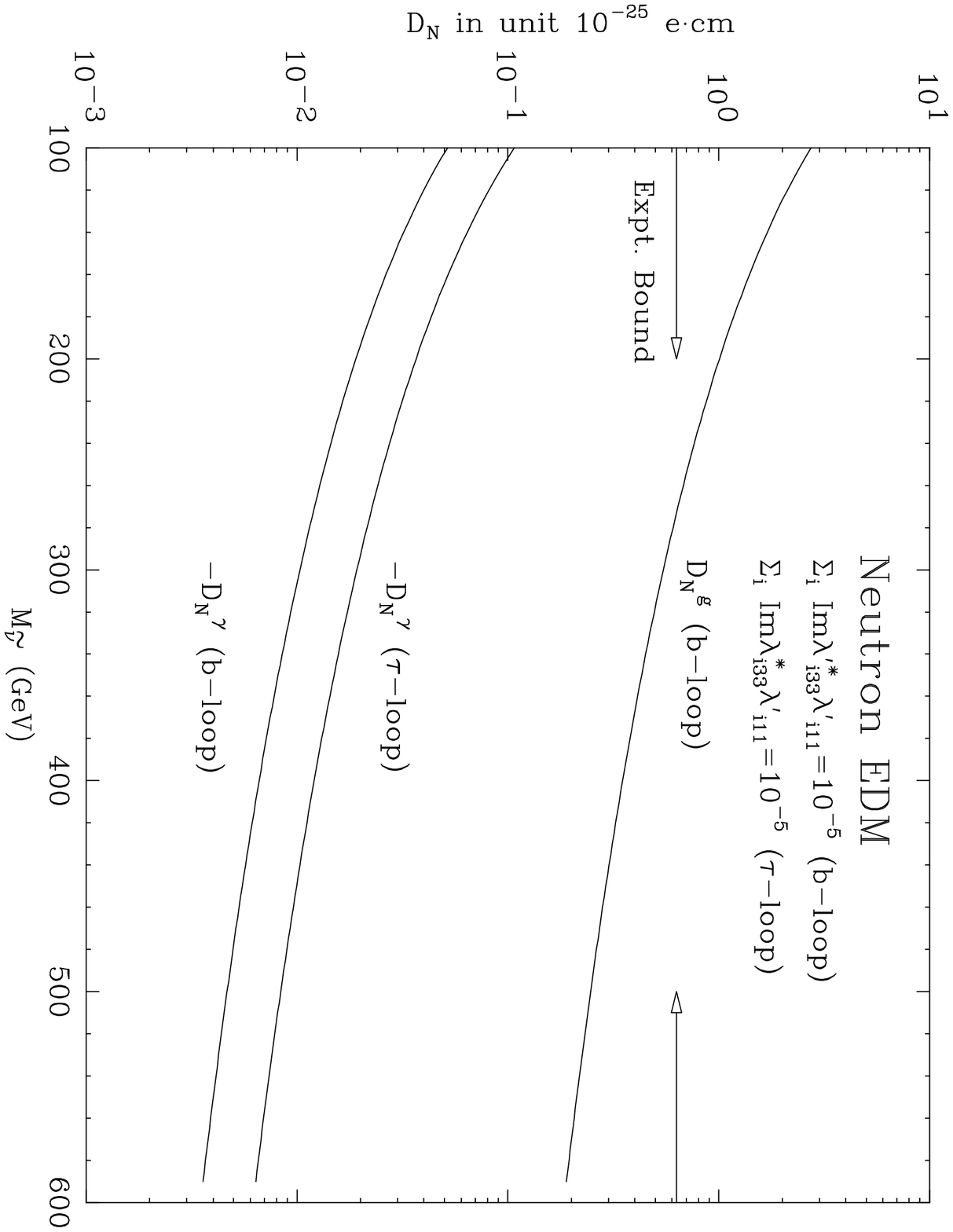}
\end{picture}\\
{\small Fig.~3
The neutron EDM $D_n$
versus $M_{\tilde \nu}$  with
$\sum_i \mbox{Im}(\lambda^{\prime * }_{i33} \lambda^{\prime}_{i11})$
or
$\sum_i \mbox{Im}(\lambda^{ * }_{i33} \lambda^{\prime}_{i11})$
scaled to $10^{-5}$.
}
\end{center}

The electron EDM can arise via both  $\lambda^{\prime}$ or $\lambda$-type
$R$-parity violating coupling
Ref.\cite{godbole}. Based on above study, the analytical
formula for the electron EDM at the two-loop level is:
$$ \left({d_e^\gamma \over e}\right) \ = -{ \alpha_{\rm em}   \over 16 \pi^3}
     \left[ \sum_{i\ne1} {3e_b^2 m_b \over  M_{\tilde \nu_i}^2   }
      \mbox{Im}(\lambda^{\prime * }_{i33}\lambda_{i11} )
     \cdot F\left({m_b^2\over M_{\tilde \nu_i}^2}\right) \right. $$
\begin{equation}
\qquad\qquad\qquad\qquad\qquad +\left.{m_\tau \over M_{\tilde
\nu_2}^2}
      \mbox{Im}(\lambda^{*}_{233}\lambda_{211} )
     \cdot F\left({m_\tau^2\over M_{\tilde \nu_2}^2}\right) \right]
    \quad .
\label{eq:edm}\end{equation}
In Fig.~4 we assume all $M_{\tilde \nu_i}$ to be equal and plot
contributions to the electron EDM versus the sneutrino mass
$M_{\tilde \nu}$ in the region of interest ($100$ to $600$ GeV).
Using the up-to-dated  experimental \cite{electronedm} bound
$|d_e|<0.43 \times 10^{-26}$ $e$-cm
and barring from accidental cancellation among contributions,
we derive   constraints:
\begin{eqnarray}
\mbox{Im}(\lambda^{*}_{233} \lambda_{211})
          &<&      0.74 \times 10^{-5} \ ,  \\
\sum_{i\ne 1} \mbox{Im}(\lambda^{\prime *}_{i33} \lambda_{i11})
          &<&      1.3  \times 10^{-5} \ , \end{eqnarray}
for $M_{\tilde \nu}=300$ GeV.

\begin{center}
\begin{picture}(400,240)(0,0)
\includegraphics{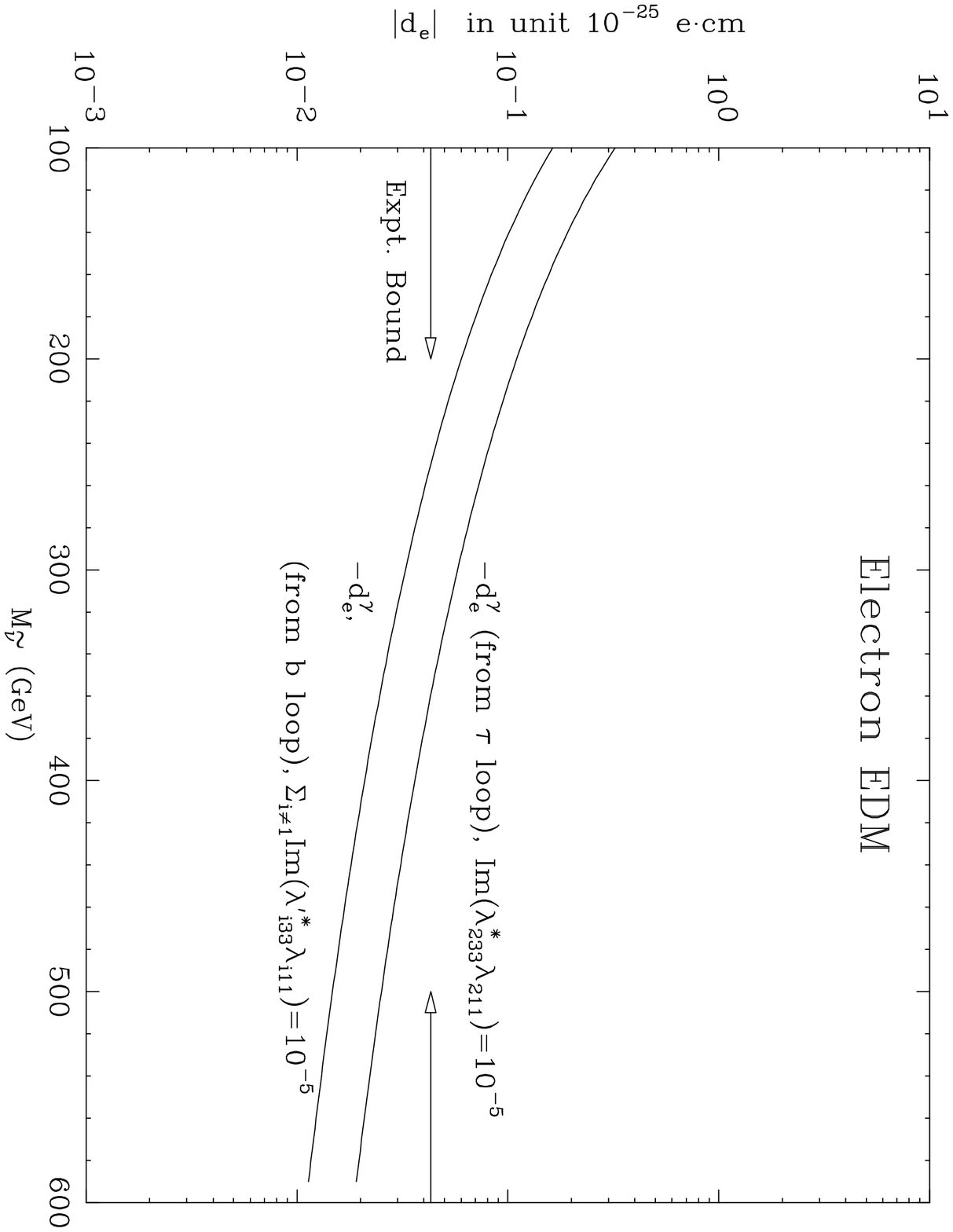}
\end{picture}\\
{\small Fig.~4
The electron EDM $d_e$
versus $M_{\tilde \nu}$.
}
\end{center}

\section*{Conclusion}
We have presented an exact and complete calculation of the
dominant contribution to the neutron EDM in a minimal supersymmetric model
without $R$ parity due to the couplings $\lambda$ and $\lambda'$.  The
$CP$ violation does not depend on the complex
phases $\phi_{\mu}$ and $\phi_{A_0}$ (the phases of the
Higgsino mass parameter and the trilinear scalar coupling $A_0$)
in Minimal Supergravity models,
therefore unrelated to the restrictive bounds or complicated cancellations
necessary in MSSM. The leading $\not\! R$ contribution to the neutron EDM
occurs at two-loop level through the Barr-Zee mechanism. We obtain
stringent bounds on the product  $\mbox{Im}\lambda^{\prime \ast
}_{i33} \lambda^{\prime}_{i11} <{\cal O}(10^{-5})$.

\noindent {\bf Acknowledgments} This work was supported in parts
by National Science Council of R.O.C., by U.S. Department of
Energy (Grant No. DE-FG02-84ER40173) and by NSERC of Canada (Grant
No. SAP0105354). MF and WFC would like to thank the High Energy
Physics Group at the University of Illinois at Chicago for their
hospitality while this work was initiated.

\bibliographystyle{unsrt}

\newpage
\section*{Figure Captions}
\begin{itemize}
\item[Fig.~1\ ]
\begin{itemize}
\item[a.]  
(i)  Rainbow-like  diagram for the $d$ quark. The generic
$\not\! R$ vertex is marked by $\circ$  and its complex conjugate by
$\bullet$.
(ii)   Rainbow-like  diagram for the $u$ quark.
\item[b.] 
(i) Overlapping  diagram for the $d$ quark using
$\lambda'$.
(ii) Overlapping diagram for the $u$ quark using
$\lambda'$.
\item[c.]
(i) Tent-like  diagram for the $d$ quark using $\lambda'$.
(ii) Tent-like  diagram for the $u$ quark using
$\lambda'$.
\item[d.]
Barr-Zee type graph for $u$ quark EDM
\end{itemize}
\item[Fig.~2\ ]
A typical two-loop diagram of the Barr-Zee type.
Note that there are 3 ways to insert mass.
\item[Fig.~3\ ]
The neutron EDM $D_n$
versus $M_{\tilde \nu}$  with
$\sum_i \mbox{Im}(\lambda^{\prime * }_{i33} \lambda^{\prime}_{i11})$
or
$\sum_i \mbox{Im}(\lambda^{ * }_{i33} \lambda^{\prime}_{i11})$
scaled to $10^{-5}$.
\item[Fig.~4\ ]
The electron EDM $d_e$
versus $M_{\tilde \nu}$.
\end{itemize}
\end{document}